\documentclass[twocolumn,aps,prb,amsmath,floatfix]{revtex4}
\usepackage{graphicx}
\begin{document}
\title{Comment on ``Absence of spin liquid in non-frustrated correlated systems''}
\author{Ansgar Liebsch}
\affiliation{Peter Gr\"unberg Institut, Forschungszentrum J\"ulich, 
             52425 J\"ulich, Germany}
\begin{abstract} %
\end{abstract}
\maketitle

In a recent Letter, Hassan and S\'en\'echal [1] discussed the existence of 
a spin-liquid phase of the half-filled Hubbard model on the honeycomb lattice. 
Using schemes, such as the variational cluster approximation (VCA) and the cluster 
dynamical mean field theory (CDMFT) in combination with exact diagonalization 
(ED), they argued that a single bath orbital per site of the six-atom unit cell 
is insufficient and leads to the erroneous conclusion that the system is gapped 
for all nonzero values of the onsite Coulomb interaction $U$. In contrast, we 
point out here that, in the case of the honeycomb lattice, six bath levels 
per six-site unit cell are perfectly adequate for the description of short-range 
correlations. Instead, we demonstrate that it is the violation of long-range 
translation symmetry inherent in CDMFT-like schemes which opens a gap at
Dirac points. The gap found at small $U$ therefore does not correspond to a 
Mott gap. As a result, present CDMFT schemes are not suitable for the 
identification of a spin-liquid phase on the honeycomb lattice.
     
As shown in Ref.~[2], the cluster self-energy obtained within ED CDMFT [3] 
using six bath levels is in nearly quantitative agreement with results 
derived within continuous-time quantum Monte Carlo (QMC) CDMFT [4]. As the 
bath in QMC is infinite, the self-energy is not subject to finite-size effects. 
The reason for the good agreement is that, because of the semi-metallic nature 
of the honeycomb lattice, the projection of the infinite-lattice bath Green's 
function onto a finite-cluster Anderson Green's function is not plagued by the  
low-energy-low-temperature disparities that arise in the case of correlated metals
and the square-lattice Hubbard model. 
Moreover, because of the hexagonal symmetry, this projection can be performed 
in the diagonal molecular-orbital basis with non-symmetrical density of states 
components [3]. The main issue in Ref.~[1] concerning the symmetry of bath levels 
then does not arise and the two independent bath Green's function components are 
fitted accurately using a total of six parameters (see Fig. 24 of Ref. [2]), 
while in the site basis only one fit parameter is available.
      
Within CDMFT, the self-energy at Dirac points $K$ exhibits the low-energy behavior [3]:
$\Sigma(K,i\omega_n) \approx a i\omega_n + b^2/[i\omega_n(1-a)]$, where $a$ is 
the initial slope of Im\,$[\Sigma_{11}(i\omega_n)-\Sigma_{13}(i\omega_n)]$ and 
$b$ the low-energy limit of Re\,$[\Sigma_{12}(i\omega_n)-\Sigma_{14}(i\omega_n)]$.
Here, $\Sigma_{ij}$ is the self-energy in the site basis and $\omega_n$ are 
Matsubara frequencies. 
The excitation gap at temperatures $T \rightarrow 0$ is given by 
$\Delta\approx 2\vert b\vert/\sqrt{1-a}$ (see also Ref.~[5]). Using the
site notation: ${\bf a}_1=(0,0)$, ${\bf a}_2=(0,1)$, ${\bf a}_3=(\sqrt3/2,3/2)$, 
${\bf a}_4=(\sqrt3,1)$, ${\bf a}_5=(\sqrt3,0)$, and ${\bf a}_6=(\sqrt3/2,-1/2)$, 
the elements $\Sigma_{12}$ and $\Sigma_{14}$ within this cell are independent, 
so that $\Delta\ne0$. However, site 1 is also connected to site 4 at 
$(-\sqrt3/2,-1/2)$ in the neighboring cell, requiring $\Sigma_{12}=\Sigma_{14}$
which is not obeyed in CDMFT. Therefore, this violation of translation symmetry is 
responsible for the insulating contribution $\sim 1/i\omega_n$ to the self-energy 
at $K$. Clearly, this term is not caused by the finite size nor the symmetry 
properties of the bath employed in ED. In fact, in view of the good agreement with 
ED, the self-energy in QMC CDMFT [4] exhibits the same insulating contribution, 
so that the density of states should also reveal a gap at small $U$ and low $T$. 

To restore translation symmetry, we have performed ED calculations within the 
dynamical cluster approximation (DCA). This approach ensures $\Sigma_{12}=\Sigma_{14}$, 
giving semi-metallic rather than insulating behavior at small $U$ [6]. The condition 
$\Sigma_{12}=\Sigma_{14}$, however, cannot be generally correct for correlations 
within the unit cell. The critical Coulomb interaction $U_c\approx 6$ found in DCA 
should therefore be used with caution.     

For interactions in the range $U \approx 4\dots 5$, ED and QMC CDMFT [3,4] yield 
excitation gaps which agree remarkably well with the charge gap derived in 
large-scale QMC calculations [7]. Evidently, this gap corresponds to a Mott gap 
induced by short-range correlations and is only weakly affected by the lack of 
long-range translation symmetry. On the other hand, the key question as to how this 
gap closes, i.e., how the semi-metallic phase with weakly distorted Dirac cones is 
recovered at small $U$, cannot be studied adequately within CDMFT. The gap obtained at 
small $U$ is an artifact caused by the violation of long-range lattice symmetry and does 
not represent a true Mott gap. As a result, the identification of a spin-liquid 
phase on the honeycomb lattice is presently not feasible within CDMFT-like methods.      
       
In conclusion, the origin of the narrow excitation gap at small $U$ found in 
ED CDMFT for the honeycomb lattice is not the finite bath as stated in Ref. [1] 
but the lack of translation symmetry. This problem is also shared by QMC CDMFT 
and can be avoided, for instance, within ED/QMC DCA.

\bigskip
\parindent=0pt
[1] S.R. Hassan and D. S\'en\'echal,  
    Phys. Rev. Lett. {\bf 110},\\ \mbox{\hskip4mm} 096402 (2013).

[2] A. Liebsch and H. Ishida, 
    J. Phys. Condensed Matter\\ \mbox{\hskip4mm} {\bf 24}, 053201 (2012).

[3] A. Liebsch, 
    Phys. Rev. B {\bf 83}, 035113 (2011).

[4] W. Wu, Y.H. Chen, H.Sh. Tao, N.H. Tong, and W.M.\\ \mbox{\hskip4mm} Liu,
    Phys. Rev. B {\bf 82}, 245102 (2010).

[5] K. Seki and Y. Ohta, 
    arXiv:1209.2101.

[6] A. Liebsch, to be published.



[7] Z.Y. Meng, T.C. Lang, S. Wessel, F.F. Assaad, and\\ \mbox{\hskip4mm} A. Muramatsu,
    Nature {\bf 464}, 847 (2010).


\end{document}